\input harvmac
\input amssym
\let\includefigures=\iftrue
\newfam\black
\input rotate
\input epsf
\noblackbox
\includefigures
\message{If you do not have epsf.tex (to include figures),} \message{change the
option at the top of the tex file.}
\def\figin{\epsfcheck\figin}\def\figins{\epsfcheck\figins}
\def\epsfcheck{\ifx\epsfbox\UnDeFiNeD
\message{(NO epsf.tex, FIGURES WILL BE IGNORED)}
\gdef\figin##1{\vskip2in}\gdef\figins##1{\hskip.5in}
\else\message{(FIGURES WILL BE INCLUDED)}%
\gdef\figin##1{##1}\gdef\figins##1{##1}\fi}
\def\DefWarn#1{}

\def\figinsert{\goodbreak\midinsert}
\def\ifig#1#2#3{\DefWarn#1\xdef#1{fig.~\the\figno}
\writedef{#1\leftbracket fig.\noexpand~\the\figno}%
\figinsert\figin{\centerline{#3}}\medskip\centerline{\vbox{\baselineskip12pt
\advance\hsize by -1truein\noindent\footnotefont{\bf Fig.~\the\figno:} #2}}
\bigskip\endinsert\global\advance\figno by1}
\else
\def\ifig#1#2#3{\xdef#1{fig.~\the\figno}
\writedef{#1\leftbracket fig.\noexpand~\the\figno}%
\global\advance\figno by1} \fi


\lref\AD{S. Ashok and M. Douglas, ``Counting Flux Vacua,''
JHEP {\bf 0401} (2004) 060,
hep-th/0307049.}

\lref\douglas{M. Douglas, ``The Statistics of String/M-theory
Vacua,'' JHEP {\bf 0305} (2003) 046, hep-th/0303194.}

\lref\dougmath{M. Douglas, B. Shiffman and S. Zelditch, ``Critical
Points and Supersymmetric Vacua,'' arXiv:math.CV/0402326.}

\lref\KT{ A.~Klemm and S.~Theisen,
``Considerations of One-Modulus Calabi-Yau Compactifications:
Picard-Fuchs Equations, K\"ahler Potentials and
Mirror Maps,'' Nucl.Phys. {\bf B389} (1993) 153, hep-th/9205041.}

\lref\GKP{S. Giddings, S. Kachru, and J. Polchinski, ``Hierarchies
from Fluxes in String Compactifications,'' Phys. Rev. {\bf D66}
(2002) 106006, hep-th/0105097.}

\lref\dG{O. DeWolfe and S. Giddings, ``Scales and hierarchies in
warped compactifications and brane worlds,'' Phys. Rev. {\bf D67} (2003)
066008, hep-th/0208123.}

\lref\IINK{
S. Gurrieri, J. Louis, A. Micu and D. Waldram, ``Mirror Symmetry
in Generalized Calabi-Yau Compactifications,'' Nucl. Phys. {\bf B654}
(2003) 61, hep-th/0211102\semi
S. Kachru, M. Schulz, P. Tripathy and S. Trivedi, ``New Supersymmetric
String Compactifications,'' JHEP {\bf 0303} (2003) 061, hep-th/0211182.}

\lref\CHSW{P. Candelas, G. Horowitz, A. Strominger and E. Witten,
``Vacuum Configurations for Superstrings,'' Nucl. Phys. {\bf B258} (1985) 46.}

\lref\unp{S. Kachru, unpublished; E. Silverstein, unpublished; L. Susskind, unpublished.}

\lref\ddf{F. Denef, M. Douglas and B. Florea, ``Building a better
racetrack,'' hep-th/0404257.}

\lref\KST{S. Kachru,~M. Schulz and S. P. Trivedi,
``Moduli Stabilization from Fluxes in a Simple IIB Orientifold,''
JHEP {\bf 0310} (2003) 007, hep-th/0201028\semi
A. Frey and J. Polchinski, ``N=3 Warped Compactifications,'' Phys. Rev.
{\bf D65} (2002) 126009, hep-th/0201029.}

\lref\GP{B. Greene and R. Plesser, ``Duality in Calabi-Yau Moduli Space,''
Nucl. Phys. {\bf B338} (1990) 15.}

\lref\BP{R. Bousso and J. Polchinski, ``Quantization of Four-Form Fluxes and
Dynamical Neutralization of the Cosmological Constant,'' JHEP {\bf 0006} (2000) 006,
hep-th/0004134.}

\lref\cdgp{P. Candelas, X. de la Ossa, P. Green and L. Parkes, ``A Pair of
Calabi-Yau Manifolds as an Exactly Soluble Superconformal Theory,''
Nucl. Phys. {\bf B359} (1991) 21.}

\lref\TT{P. Tripathy and S. P. Trivedi, ``Compactification with Flux
on K3 and Tori,''  JHEP {\bf 0303} (2003) 028, hep-th/0301139.}

\lref\GKTT{A. Giryavets, S. Kachru, P. Tripathy and S. Trivedi,
``Flux Compactifications on Calabi-Yau Threefolds,''
JHEP {\bf 0404} (2004) 003, hep-th/0312104.}

\lref\KKLT{S. Kachru, R. Kallosh, A. Linde and S.P. Trivedi, ``de Sitter Vacua
in String Theory,'' Phys. Rev. {\bf D68} (2003) 046005, hep-th/0301240.
}

\lref\denef{F. Denef and M. Douglas, ``Distributions of Flux Vacua,''
hep-th/0404116.}

\lref\dtalk{M. Douglas, various talks, Fall 2003.}

\lref\feng{J. Feng, J. March-Russell, S. Sethi and F. Wilczek,
``Saltatory Relaxation of the Cosmological Constant,''
Nucl. Phys. {\bf B602} (2001) 307, hep-th/0005276.}

\lref\Herman{H. Verlinde, ``Holography and Compactification,''
Nucl. Phys. {\bf B580} (2000) 264, hep-th/9906182\semi
C. Chan, P. Paul and H. Verlinde, ``A Note on Warped
String Compactification,'' Nucl. Phys. {\bf B581} (2000) 156,
hep-th/0003236.}

\lref\OtherIIB{
A. Saltman and E. Silverstein, ``The Scaling of the No-Scale Potential
and de Sitter Model Building,'' hep-th/0402135\semi
C.P. Burgess, R. Kallosh and F. Quevedo, ``De Sitter Vacua from Supersymmetric
D-terms,'' JHEP {\bf 0310} (2003) 056, hep-th/0309187\semi
C. Escoda, M. Gomez-Reino and F. Quevedo, ``Saltatory de Sitter String
Vacua,'' JHEP {\bf 0311} (2003) 065, hep-th/0307160\semi
A. Frey, M. Lippert and B. Williams, ``The Fall of Stringy
de Sitter,'' Phys. Rev. {\bf D68} (2003) 046008, hep-th/0305018.
}

\lref\KPV{
S. Kachru, J. Pearson and H. Verlinde, ``Brane/Flux Annihilation and
the String Dual of a Non-supersymmetric Field Theory,'' JHEP
{\bf 0206} (2002) 021, hep-th/0112197.}

\lref\nonkahler{For interesting work in this direction with further
references, see:
G.L. Cardoso, G. Curio, G. Dall'Agata, D. L\"ust, P. Manousselis
and G. Zoupanos, ``Nonkahler String Backgrounds and their Five
Torsion Classes,'' Nucl. Phys. {\bf B652} (2003) 5, hep-th/0211118\semi
K. Becker, M. Becker, K. Dasgupta and P. Green, ``Compactifications
of Heterotic Theory on NonK\"ahler Complex Manifolds I,''
JHEP {\bf 0304} (2003) 007, hep-th/0301161\semi
K. Becker, M. Becker, K. Dasgupta, P. Green,
E. Sharpe, ``Compactifications of Heterotic Strings on NonK\"ahler
Complex Manifolds II," Nucl. Phys. {\bf B678} (2004) 19, hep-th/0310058\semi
J. Gauntlett, D. Martelli and D. Waldram, ``Superstrings with
Intrinsic Torsion,'' Phys. Rev. {\bf D69} (2004) 086002, hep-th/0302158.
}
\lref\ps{J. Polchinski and M. Strassler, ``The String Dual of a
Confining Four-Dimensional Gauge Theory,'' hep-th/0003136.}

\lref\RS{L. Randall and R. Sundrum, `` A Large Mass Hierarchy from
a Small Extra Dimension,'' Phys. Rev. Lett. {\bf 83} (1999) 3370,
hep-th/9905221.}

\lref\GVW{S. Gukov, C. Vafa and E. Witten, ``CFTs from Calabi-Yau Fourfolds,''
Nucl. Phys. {\bf B584} (2000) 69, hep-th/9906070\semi
T.R. Taylor and C. Vafa, ``RR flux on Calabi-Yau and partial supersymmetry
breaking,'' Phys. Lett. {\bf B474} (2000) 130, hep-th/9912152\semi
P. Mayr, ``On Supersymmetry Breaking in String Theory and its Realization
in Brane Worlds,'' Nucl. Phys. {\bf B593} (2001) 99, hep-th/0003198.}

\lref\MSS{A. Maloney, E. Silverstein and A. Strominger, ``De Sitter Space
in Noncritical String Theory,'' hep-th/0205316.}

\lref\lerche{W. Lerche, ``Special Geometry and Mirror Symmetry for Open
String Backgrounds with ${\cal N}=1$ Supersymmetry,'' hep-th/0312326.}

\lref\Heterotic{
M. Becker, G. Curio and A. Krause, ``De Sitter Vacua from
Heterotic M Theory,'' hep-th/0403027\semi
R. Brustein and S.P. de Alwis, ``Moduli Potentials in String
Compactifications with Fluxes: Mapping the Discretuum,'' hep-th/0402088\semi
S. Gukov, S. Kachru, X. Liu and L. McAllister,
``Heterotic Moduli Stabilization with Fractional Chern-Simons Invariants,''
hep-th/0310159\semi
E. Buchbinder and B. Ovrut, ``Vacuum Stability in Heterotic M-theory,''
hep-th/0310112.}

\lref\svw{S. Sethi, C. Vafa and E. Witten, ``Constraints on Low-Dimensional
String Compactifications,'' Nucl. Phys. {\bf B480} (1996) 213, hep-th/9606122.}

\lref\oldflux{
A. Strominger, ``Superstrings with Torsion,'' Nucl. Phys. {\bf B274} (1986) 253\semi
J. Polchinski and A. Strominger, ``New vacua for
type II string theory,'' Phys. Lett. {\bf B388} (1996) 736,
hep-th/9510227\semi K. Becker and M. Becker, ``M-theory on eight
manifolds,'' Nucl. Phys. {\bf B477} (1996) 155,
hep-th/9605053\semi J.Michelson, ``Compactifications of type IIB
strings to four-dimensions with non-trivial classical potential,''
Nucl. Phys. {\bf B495} (1997) 127, hep-th/9610151\semi
K. Dasgupta, G. Rajesh and S. Sethi, ``M-theory, orientifolds and G-flux,''
JHEP {\bf 9908} (1999) 023, hep-th/9908088\semi
B. Greene, K. Schalm and G. Shiu, ``Warped compactifications in M and F theory,''
Nucl. Phys. {\bf B584} (2000) 480, hep-th/0004103\semi
G. Curio, A. Klemm, D. L\"ust and S. Theisen, ``On the Vacuum Structure of
Type II String Compactifications on Calabi-Yau Spaces with H-Fluxes,''
Nucl. Phys. {\bf B609} (2001) 3, hep-th/0012213\semi
K. Becker and M. Becker, ``Supersymmetry Breaking, M Theory and
Fluxes,'' JHEP {\bf 010} (2001) 038, hep-th/0107044\semi
M. Haack and J. Louis, ``M theory compactified on Calabi-Yau
fourfolds with background flux,'' Phys. Lett. {\bf B507} (2001)
296, hep-th/0103068\semi J. Louis and A. Micu, ``Type II theories
compactified on Calabi-Yau threefolds in the presence of
background fluxes,'' Nucl.Phys. {\bf B635} (2002) 395, hep-th/0202168.}

\lref\ferrara{R. D'Auria, S. Ferrara and S. Vaula, ``${\cal N}=4$
gauged supergravity and a IIB orientifold with fluxes,'' New J.
Phys. {\bf 4} (2002) 71, hep-th/0206241\semi S. Ferrara and M.
Porrati, ``${\cal N}=1$ no-scale supergravity from IIB
orientifolds,'' Phys. Lett. {\bf B545} (2002) 411,
hep-th/0207135\semi R. D'Auria, S. Ferrara, M. Lledo and S. Vaula,
``No-scale ${\cal N}=4$ supergravity coupled to Yang-Mills: the
scalar potential and super-Higgs effect,'' Phys. Lett. {\bf B557}
(2003) 278, hep-th/0211027\semi R. D'Auria, S. Ferrara, F.
Gargiulo, M. Trigiante and S. Vaula, ``${\cal N}=4$ supergravity
Lagrangian for type IIB on $T^6/Z_2$ orientifold in presence of
fluxes and D3 branes,'' JHEP {\bf 0306} (2003) 045,
hep-th/0303049\semi L. Andrianopoli, S. Ferrara and M. Trigiante,
``Fluxes, supersymmetry breaking and gauged supergravity,''
hep-th/0307139\semi B. de Wit, Henning Samtleben and M. Trigiante,
``Maximal Supergravity from IIB Flux Compactifications,''
hep-th/0311224.}

\lref\sen{A. Sen, ``Orientifold limit of F-theory vacua,'' Phys. Rev.
{\bf D55} (1997) 7345, hep-th/9702165.}

\lref\softsusy{P.G. Camara, L.E. Ibanez and A.M Uranga, ``Flux Induced
SUSY Breaking Soft Terms, hep-th/0311241\semi
M. Grana, T. Grimm, H. Jockers and J. Louis, ``Soft Supersymmetry
Breaking in Calabi-Yau Orientifolds with D-branes and Fluxes,''
hep-th/0312232\semi
A. Lawrence and J. McGreevy, ``Local String Models of Soft Supersymmetry
Breaking,'' hep-th/0401034.}

\lref\susskind{L. Susskind, ``The anthropic landscape of string theory,''
hep-th/0302219.}

\lref\Moore{G. Moore, ``Les Houches Lectures on Strings and Arithmetic,''
hep-th/0401049.}

\lref\KS{I. Klebanov and M. Strassler,
``Supergravity and a Confining Gauge Theory: Duality Cascades and $\chi$SB
Resolution of Naked Singularities,'' JHEP {\bf 008} (2000) 052, hep-th/0007191.}

\lref\uranga{ J. Cascales, M. Garcia de Moral, F. Quevedo and A.
Uranga, ``Realistic D-brane Models on Warped Throats: Fluxes,
Hierarchies and Moduli Stabilization,'' hep-th/0312051\semi J.
Cascales and A. Uranga, ``Chiral 4d String Vacua with D-branes and
Moduli Stabilization,'' hep-th/0311250\semi J. Cascales and A.
Uranga, ``Chiral 4d ${\cal N}=1$ String Vacua with D-Branes and
NSNS and RR Fluxes,'' JHEP {\bf 0305} (2003) 011,
hep-th/0303024\semi
R. Blumenhagen, D. L\"ust and T. Taylor, ``Moduli Stabilization in Chiral
Type IIB Orientifold Models with Fluxes,'' Nucl. Phys. {\bf B663} (2003) 219,
hep-th/0303016.
}

\lref\BDG{T. Banks, M. Dine and E. Gorbatov,
``Is there a string theory landscape?,'' hep-th/0309170.}
\lref\Susskind{L. Susskind, ``The Anthropic Landscape of String Theory,''
hep-th/0302219.}


\def\IL{\relax{\rm I\kern-.18em L}}
\def\IH{\relax{\rm I\kern-.18em H}}
\def\IR{\relax{\rm I\kern-.18em R}}
\def\IC{\relax\hbox{$\inbar\kern-.3em{\rm C}$}}
\def\IZ{\relax\ifmmode\mathchoice
{\hbox{\cmss Z\kern-.4em Z}}{\hbox{\cmss Z\kern-.4em Z}}
{\lower.9pt\hbox{\cmsss Z\kern-.4em Z}} {\lower1.2pt\hbox{\cmsss Z\kern-.4em
Z}}\else{\cmss Z\kern-.4em Z}\fi}
\def\CM {{\cal M}}

\def\CG {{\cal G}}

\def\CM {{\cal M}}

\font\manual=manfnt \def\dbend{\lower3.5pt\hbox{\manual\char127}}

\def\IZ{\relax\ifmmode\mathchoice
{\hbox{\cmss Z\kern-.4em Z}}{\hbox{\cmss Z\kern-.4em Z}}
{\lower.9pt\hbox{\cmsss Z\kern-.4em Z}} {\lower1.2pt\hbox{\cmsss Z\kern-.4em
Z}}\else{\cmss Z\kern-.4em Z}\fi}

\chardef\tempcat=\the\catcode`\@ \catcode`\@=11
\def\cyracc{\def\u##1{\if \i##1\accent"24 i%
    \else \accent"24 ##1\fi }}
\newfam\cyrfam
\font\tencyr=wncyr10
\def\cyr{\fam\cyrfam\tencyr\cyracc}

\def\bar{\overline}

\def\rt2{\sqrt{2}}
\def\irt2{{1\over\sqrt{2}}}

\def\slashchar#1{\setbox0=\hbox{$#1$}           
   \dimen0=\wd0                                 
   \setbox1=\hbox{/} \dimen1=\wd1               
   \ifdim\dimen0>\dimen1                        
      \rlap{\hbox to \dimen0{\hfil/\hfil}}      
      #1                                        
   \else                                        
      \rlap{\hbox to \dimen1{\hfil$#1$\hfil}}   
      /                                         
   \fi}
\writedefs

%
%
%
\newbox\tmpbox\setbox\tmpbox\hbox{\abstractfont }

\Title{\vbox{\baselineskip12pt\hbox{hep-th/0404243}
\hbox{HUTP-04/A019, SU-ITP-04/16, SLAC-PUB-10425, TIFR/TH/04-11} }}
 {\vbox{ {\centerline{On the Taxonomy of Flux Vacua}
}}}
\centerline{Alexander
Giryavets$^a$\footnote{*}{On leave from Steklov Mathematical
Institute, Moscow, Russia}\footnote{$^1$}{giryav@stanford.edu}, Shamit
Kachru$^a$\footnote{$^2$}{skachru@stanford.edu} and
Prasanta K. Tripathy$^{b,c}$\footnote{$^3$}{prasanta@theory.tifr.res.in}}
\smallskip
\centerline{$^{a}$ Department of Physics and SLAC}
\centerline{Stanford University} \centerline{Stanford, CA
94305/94309 USA}
\smallskip
\centerline{$^{b}$ Tata Institute for Fundamental Research}
\centerline{Homi Bhabha Road, Mumbai 400 005, INDIA}
\smallskip
\centerline{$^{c}$ Jefferson Laboratory}
\centerline{Harvard University}
\centerline{Cambridge, MA 02138 USA}

\bigskip

\noindent
We investigate several predictions about the properties of IIB flux vacua on
Calabi-Yau orientifolds, by constructing and characterizing a very
large set of vacua in a specific example, an orientifold of the
Calabi-Yau hypersurface in $WP^{4}_{1,1,1,1,4}$.   We find support
for the prediction of Ashok and Douglas that the density of vacua on
moduli space is governed by ${\rm det}(-R - \omega)$ where $R$ and $\omega$
are curvature and K\"ahler forms on the moduli space.
The conifold point $\psi=1$ on moduli space therefore serves as an attractor,
with a significant fraction of the flux vacua contained in a small neighborhood
surrounding $\psi=1$.
We also
study the functional dependence of the number of flux vacua on the
D3 charge in the fluxes, finding simple power law growth.

\Date{April 2004}

\newsec{Introduction}

Since the classic work of Candelas, Horowitz, Strominger and Witten in the mid 1980s
\CHSW, Calabi-Yau compactification of 10d supergravity has been the dominant paradigm
for the construction of 4d string models.  In the heterotic string theory,
such compactifications
yield ${\cal N}=1$ supersymmetric
theories below the KK scale, and can quite plausibly accomodate
the standard model.
In the type II context, Calabi-Yau compactification yields ${\cal N}=2$ supersymmetric
models, which have been of considerable interest in studies of string duality
and mirror symmetry.

More recently, it has become clear that these models are only the simplest representatives
of a much larger class of models.  For concreteness, we will restrict our discussion
to the type IIB theory, but
similar remarks would
apply in the other limits of M-theory.
In the IIB context, the theory contains p-form gauge fields which couple to the various
supersymmetric D-branes and NS-branes.  One can consider magnetic
fluxes of these gauge fields
in the internal dimensions; such fluxes are consistent with 4d Lorentz invariance.
In fact, in type II orientifold models which break the supersymmetry from ${\cal N}=2$
to ${\cal N}=1$, generic methods of satisfying the tadpole conditions will involve
the addition of background magnetic flux.
This is interesting because the magnetic fluxes will cost an energy which depends
on the detailed shape and size of the internal dimensions; that is, they will generate
a potential for some of the Calabi-Yau moduli.

In the case of type IIB on a Calabi-Yau orientifold, the allowed fluxes
are those of the RR and NS three-form field strengths $F_3$ and $H_3$.  The properties
of the theory in the presence of such fluxes have been investigated in a vast literature;
see for instance \refs{\oldflux \GVW \GKP \KST \TT{--} \GKTT}.
The resulting
${\cal N}=1$ supergravity theories (derived in \refs{\GKP,\dG}) are of the no-scale type,
with a K\"ahler potential $K$
inherited (at large volume and weak coupling) from the Calabi-Yau compactification
and a superpotential which is computable in terms of the choice of quantized
background magnetic fluxes \GVW.
The resulting scalar potential
generically fixes the complex structure moduli and the dilaton.
It is possible to make semi-realistic chiral models of particle physics in this
framework \uranga, and to describe some aspects of the resulting supergravity
theories in the general
framework of gauged extended supergravity \ferrara.
The theory of soft susy breaking terms in these models has been developed
in \softsusy.

For a given choice of the integral background fluxes,
the superpotential for moduli can simply be written as a
combination of periods of the holomorphic three-form $\Omega$ on the Calabi-Yau space.
The periods are computable by standard techniques in classical geometry, highly
developed in the physics literature after the seminal work of Candelas et al on
mirror symmetry \cdgp.  The resulting superpotential is exact at least to all orders
in the $\alpha^\prime$ expansion by standard holomorphy arguments.
It can be argued that ``generic'' non-perturbative effects (known to occur in various
models in various circumstances, and likely to occur in a generic setting due to
the absence of symmetries that would prevent them) should
suffice to fix the remaining (K\"ahler)
moduli in some cases \refs{\KKLT,\douglas}.
As argued in \KKLT, if the SUSY-breaking ($e^K|W|^2$ evaluated in vacuum)
by the fluxes is suitably small, and if
$g_s$ is stabilized by the fluxes at weak coupling, one can have small
expansion parameters which allow one to stabilize the remaining moduli
in a regime of control.  Of course, this will only happen in at most a
small fraction of the vacua.
As emphasized in \KKLT, following the basic ideas of Bousso and Polchinski \BP
(see also \feng),
the large number of possible integral flux choices makes
it plausible that suitable flux choices do exist, which allow one to
fix $g_s$ at weak coupling and keep the
SUSY breaking hierarchically smaller than
the string scale.

This set of issues, which goes into establishing the existence of a
``landscape'' of string/M-theory vacua (like the ``discretuum'' of \BP)
and characterizing its properties,
can and should be studied systematically
and quantitatively.  In addition to the works already described, a
persuasive discussion of this can be found in the works of
Douglas and collaborators \refs{\douglas,\AD \dougmath {--} \denef} (see also
\refs{\Susskind,\BDG} for more general comments about the landscape).
In particular, in \AD, under well motivated assumptions some theorems are proven
regarding the numbers
and distribution of flux vacua on the Calabi-Yau moduli space (really,
the moduli space of complex structures and dilaton VEVs) ${\cal M}$.
The assumption which goes into proving
these theorems is, roughly, that the superpotential
is a random section (with respect to a covariance kernel determined by
the K\"ahler potential) of the appropriate line bundle over ${\cal M}$.
This models the flux superpotentials as random combinations of periods
with respect to some measure, instead of as the precise integral combinations of
periods determined by various choices of $H_3$ and $F_3$ consistent with the
tadpole conditions.  Basically, this assumption follows from the statement that
if one allows large enough fluxes, flux quantization becomes irrelevant for
the average properties of the vacua.
The natural size of the fluxes required by tadpole conditions in IIB string
theory probably falls into this range.

Because the emerging discretuum of vacua is too complicated to allow us to usefully
examine each vacuum individually, this kind of program to find simple assumptions
which accurately characterize the space of vacua seems to us rather important.
In fact, following
\refs{\AD,\denef}, one could simply disregard the K\"ahler moduli altogether, ignore
the no-scale structure (which is certainly broken by quantum corrections),
and characterize the properties of the supergravities
that result with the given $K$ and $W$ as a function of complex structure moduli.
The resulting ensemble of supergravity theories is rich enough that it is very plausible
that some aspects of the full story (about the real space of string vacua),
are
captured by this simplified ensemble.  However, we will continue to
use the language and equations appropriate to thinking of these as
vacua which arise in the no-scale approximation to
the full IIB theory, though it is easy to translate back
and forth.

In this paper, we test the prediction of \AD\ regarding the distribution of
vacua on ${\cal M}$, by studying the vacua which result
on a subspace of the moduli space of a certain Calabi-Yau threefold, in the
presence of a small subset of the possible fluxes.
While we are looking at a very small fraction of the possible fluxes
and hence of the total number of
flux vacua on this space,
the numbers even on our subspace are easily large enough to
allow a meaningful test of the conjecture.
In addition, in these vacua there is a tadpole condition that requires
$\int H_3 \wedge F_3 = N_{\rm flux} \leq L$, where $L$ is some number fixed
by global details of the model.  The growth of the number of vacua with
$L$, fixing the number of cycles which support flux, is subject to various
conjectures along the lines of \BP.  Here, in addition to testing
the distribution of vacua on moduli space, we also test the growth of
the number of vacua as a function of $L$, and find a simple power law.
A proper understanding of the
theory which determines this power is rather complicated, as we describe
in the following; some relevant results are discussed in the
revised version of \denef.

The organization of this paper is as follows.
In \S2, we
describe the predictions we wish to test.  In \S3, we discuss
the model in which we will test them
(which is ``model A'' of \GKTT).  This model has (on the locus of interest to us)
a one-dimensional complex structure moduli space, parametrized by a complex
variable $\psi$.
In sections 4 and 5, we describe how we find solutions of the flux equations in
two regions in the moduli space, using expansions of the periods
around $\psi=0$ (the mirror Landau-Ginzburg point) and $\psi=1$ (the
conifold point).
In each of these cases,
we find sufficiently large numbers of vacua to perform meaningful tests of the
predictions.  In \S6 we present our conclusions.

We should stress that although the discussion
here is specific to type IIB vacua of the class described in \refs{\GKP}
(for which several de Sitter scenarios exist \refs{\KKLT,\OtherIIB}),
a similar landscape also emerges from careful consideration
of the various other limits of M-theory (see e.g. \Heterotic\ for discussions
in the heterotic setting, and \MSS\ for an early construction of
de Sitter vacua in noncritical string theories).
For instance in the heterotic string, one needs to consider the large class
of possible gauge bundles in each $E_8$, in addition to the possibility of
NS three-form flux and Non-K\"ahlerity \nonkahler.  This is technically
more challenging than studying the class of IIB flux vacua we considered here,
but finding similar tractable toy ensembles in that setting should be possible.
Another setting where similar equations arise is in the problem of finding
BPS black holes in ${\cal N}=2$ supergravity; the interesting connections
between that problem and the problem of finding flux vacua are described in
\Moore.

While this work was being completed, the paper \denef\ which has some overlap
with our results appeared.

\newsec{The conjectures}

We will be interested in testing two conjectures about the large class of
IIB flux vacua described in \GKP, in the no-scale approximation.  The
conjectures concern the distribution of flux vacua on the moduli space
of complex structures, and the
scaling of the number of flux vacua with the D3-charge in the fluxes.
They can be proven to hold \AD\ under suitable assumptions about the
irrelevance of flux quantization, so we are really testing whether
those assumptions hold for
simple representatives of the class of vacua described in \GKP.

\subsec{Where do the vacua lie?}

In a recent paper of Ashok and Douglas \AD, it was conjectured that the
index density of the no-scale vacua described in \GKP\ is given by an
integral

\eqn\indexd{{\cal I} \sim \int_{\cal M} {\rm det}(-R - \omega) ~.}
${\cal I}$ is normalized appropriately to the total number (or index) of
no-scale vacua.
Here $\omega$ is the K\"ahler form on the moduli space and
$R$ is the curvature two-form expressed as a Hermitian matrix
\eqn\ris{R^{l}_k = R^{l}_{i \bar j k} dz^i \wedge d\bar z^{\bar j}}
with $z^i$ the coordinates on ${\cal M}$.
The moduli space ${\cal M}$ consists of the fundamental domain of any
relevant duality groups, acting on the complex structure and axio-dilaton
Teichmuller spaces.  In our example, it will be a product of the standard
fundamental domain of $SL(2,Z)$ in the upper half-plane (for the axio-dilaton
$\tau$)
with a (complex) one-dimensional slice of a Calabi-Yau complex structure
moduli space.  As described in \refs{\AD,\denef} (see especially
\S3.1.2 of \denef), one can integrate over $\tau$-space and find
a distribution of vacua  
on just the complex structure moduli space; for one-modulus models 
one finds
\eqn\onemodi{{\cal I} \sim \int_{\cal M} c_{1}~,}
where now ${\cal M}$ is just the complex structure moduli space,
and $c_1$ is the first Chern class.

We will assume that there are no miraculous cancellations, so that the
total number of vacua scales like ${\cal I}$.
In this case, one predicts
that the number of vacua in a given region of complex structure
moduli space should scale
as in \onemodi.  
We will test this conjecture in sections 4 and 5 by
computing the numbers of vacua
lying in various regions of the moduli space (by using monte carlo
simulations to choose fluxes and then solving the flux equations), and
checking that they
are distributed like the appropriate integrals of 
the first Chern class.

\subsec{Growth of number of vacua with $L$}

The three-form fluxes in type IIB supergravity contribute to the D3-brane charge
tadpole on the compact space, via an equation of the form
\eqn\fluxch{N_{\rm flux} = {1\over (2\pi)^4 (\alpha^\prime)^2} \int_M H_3 \wedge F_3}
where tadpole cancellation requires
\eqn\tadp{N_{\rm D3} + N_{\rm flux} = L~.}
Here $L$ is some number fixed by the orientifold charge in the given model
and $N_{\rm D3}$ is the number of space-time filling D3 branes, transverse to $M$.

While it may appear from \tadp\ that this is not much of a restriction, since
the quadratic form $\int H_3 \wedge F_3$ is not positive definite, this is not
quite correct.  On solutions of the equations of motion, one finds
(as on p.4 of the first reference in \KST)
that
$F_3 \sim -*_{6} H_3$, so $N_{\rm flux}$ becomes a positive semidefinite form.
This means that in a space with $K$ three-cycles, one might heuristically
think of \tadp\ as allowing one to search for flux vacua at each integral
lattice point within
a sphere of radius $\sqrt{L}$ in flux space.
Although $H$ and $F$ together involve 2K flux integers, using the relation
between $H$ and $F$ on solutions indicates that for a fixed point in moduli
space, this will be a $K-1$ sphere (instead of a $2K-1$ sphere).
So for e.g. a $(T^2)^3$ with square two-tori, one can easily rewrite
\tadp\ (setting $N_{D3}=0$)
as the equation for a sphere in $K$ dimensions.  Then, the number
of solutions for the flux equations $N_{vacua}$, like the number
of lattice points inside this sphere, presumably scales like
\eqn\nsol{N_{vacua} \sim
L^{K/2}~.}
This would be the behavior of the number of solutions, in the simple
toroidal example above, at a point in ${\cal M}$ where solutions to the flux
equations exist.
\foot{One might ask whether one could increase the number of solutions by
including anti-D3 branes (i.e. taking $N_{D3} < 0$ in \tadp).
While one can include anti-branes in these models, the number is bounded
by the decay mechanism discussed in \KPV, and we will simply assume
$N_{D3} \geq 0$ for now.}

This is the kind of logic suggested in e.g. \BP, applied in
this context by many
workers \unp, and rigorously argued to hold in this context
in \AD.
However, the results of \AD\ contain a surprise -- the predicted scaling
of the total number of vacua is actually
\eqn\ntots{N_{vacua} \sim L^K}
instead of the $L^{K/2}$ in \nsol.  To understand this,\foot{We are grateful
to M. Douglas for several discussions about this subject.}
one must carefully consider a smoothed average of the number of solutions
per unit volume in ${\cal M}$ -- the number of solutions jumps discontinuously
as one moves in ${\cal M}$,
so one must average to get a reasonable formula.  This averaging
produces an enhancement of the result -- for large enough $L$, a given
small region of moduli space will give rise to $L^{K}$ lattice points
within flux space, instead of the naive $L^{K/2}$.  This in particular
means that if one looks at finite $L$ in some region of the moduli space,
one may find an intermediate scaling that is between the naive
prediction \nsol\ and the $L^K$ behavior \ntots\ expected from \AD.
The detailed crossover behavior determining how the scaling should
go for a given region in ${\cal M}$ at various $L$ and $K$ is a complicated
subject which is still under investigation (see the revised
version of the paper by Denef and Douglas \denef\ for further discussion).

Because of the complications described above, we simply take our
results as an experimental clue that should help in the development
of the full theory of how $N_{vacua}$ scales with $L$ and $K$, though
we expect to (and do) find a power intermediate between \nsol\ and \ntots.

\newsec{The model of interest}

We will study an orientifold of the Calabi-Yau threefold $M$ which arises as a hypersurface
in the weighted projective space $WP^{4}_{1,1,1,1,4}$
\eqn\model{4x_0^2+x_1^8+x_2^8+x_3^8+x_4^8-8\psi x_0 x_1 x_2 x_3 x_4=0.}
This threefold has $h^{1,1}(M)=1$ and $h^{2,1}(M)=149$.
As explained in \GKTT, an orientifold of $M$ arises as a limit of
the F-theory compactification
on the Calabi-Yau fourfold $X$ given by a hypersurface in $WP^{5}_{1,1,1,1,8,12}$.
The orientifold action takes $x_0 \to -x_0$ and $\psi \to
-\psi$ (in conjunction with the worldsheet parity reversal), while leaving
all other variables unchanged.
This limit of the F-theory model on $X$ is
obtained by following the general prescription given by
Sen \sen.  One of the benefits of knowing the F-theory lift is that it allows us
to easily compute the tadpole condition for compactification on the orientifold
of $M$.  The relevant condition is (see \svw\ and the third paper listed in
\oldflux)

\eqn\xis{N_{\rm D3} + N_{\rm{flux}} = {\chi(X) \over 24} = 972}
where $N_{\rm D3}$ is the number of space-filling D3 branes one inserts transverse
to $M$, and $N_{\rm{flux}}$ is the amount of D3 charge carried by the $H_3$ and
$F_3$ fluxes as given in \fluxch.

In general, one could add many deformations to the defining equation \model\
which are consistent with the orientifold action.  However,
as explained in \GKTT\
following \GP,
the equation \model\ is the most general equation consistent with a global
symmetry group ${\cal G} = Z_8^2 \times Z_2$ acting on the projective
coordinates.  If one turns on only fluxes which are consistent with preserving
this symmetry, then the resulting superpotential is guaranteed to depend on
the additional moduli only at higher orders (in ${\cal G}$ invariant combinations).
This means that if we set these moduli to zero (as done in \model),
compute the periods relevant for our subset of fluxes,  and solve
the resulting equations for $\psi$ and the axio-dilaton $\tau$, we are
guaranteed that the solutions we find will be solutions of the full theory.
The additional moduli can be consistently set to zero because of the symmetry,
and will generically be constrained by a higher order potential in the resulting
vacua.

In fact, the subset of periods which are relevant to the ${\cal G}$ invariant
fluxes coincide (up to a factor of $\vert {\cal G}\vert$) with those of the
mirror manifold to $M$, $W$, as explained in \GKTT.  These correspond to periods
on four different 3-cycles in $M$, and hence we proceed to develop the relevant
formalism for computing flux superpotentials involving only $H_3$ and $F_3$ fluxes
through these four 3-cycles.
Happily, the periods of interest to us in this example were computed in one of
the early studies of mirror symmetry, by Klemm and Theisen \KT.

In comparing to the conjecture of \S2.2, we see that our problem has $K=4$
and $L=972$.
So in the physical problem of interest, we could check the number of vacua with
flux charge $N_{\rm flux} \leq L_*$ for any $L_* \leq 972$.
(Of course formally, we are free to consider vacua using the periods from this
model with arbitrary $L$).

\subsec{Homology and Cohomology bases}

We will work with a symplectic homology basis for the subspaces of
$H_3$ of interest to us. The basis of three-cycles $A_a$ and $B^a$
($a=1,2$) and the basis for
integral cohomology $\alpha_a$ and $\beta^a$ satisfy
\eqn\homcohbasis{ \int_{A_a}\alpha_{b}=\delta^{a}_{b},~~~~~~
\int_{B^b}\beta^{a}=-\delta^{a}_{b},~~~~~~
\int_{M}\alpha_{a}\wedge\beta^{b}=\delta^{b}_{a}.} The holomorphic
three form can be represented in terms of periods in this basis as
follows: \eqn\holform{ \int_{A_a}\Omega=z^a,~~~~~~
\int_{B^a}\Omega={\cal G}_a,~~~~~~ \Omega=z^{a}\alpha_a-{\cal
G}_{a}\beta^{a}.} In addition \eqn\ombom{
\int_{M}\Omega\wedge\bar{\Omega}=\bar{z}^a{\cal G}_{a}
-z^a\bar{{\cal G}}_{a}= \bar{z}^a{\partial {\cal G} \over \partial
z^a} -z^a{\partial \overline{{\cal G}} \over \partial \bar{z}^a}
=-\Pi^{\dagger}\cdot\Sigma\cdot \Pi~.}  Here, we have introduced
the prepotential ${\cal G}(z^1,z^2)$, the period vector $\Pi$
(whose entries are the periods \holform), and the matrix
\eqn\sigma{ \Sigma=\pmatrix{ 0 & 1\cr -1 & 0 \cr}} whose entries
are two by two matrices.

\subsec{Fluxes, Superpotential and K\"ahler Potential}

The NS and RR fluxes admit the following quantization condition
\eqn\quant{ F_{(3)}=(2\pi)^2\alpha'( f_a [B_a]+f_{a+2}
[A_a]),~~~~~~~ H_{(3)}=(2\pi)^2\alpha'( h_a [B_a]+h_{a+2} [A_a]) }
with integer $f_i$ and $h_i$. Here we
used the notation $[A_a]=\alpha_a$ and
$[B_a]=\beta_{a}$. Using this notation, we  find the following
expression for $N_{\rm flux}$ \eqn\nflux{ N_{\rm flux}={1\over
(2\pi)^4(\alpha')^2}\int_{M} H_{(3)}\wedge
F_{(3)}=f^{T}\cdot\Sigma\cdot h.} The superpotential is given by
\eqn\superp{ W=\int_{M} (F_{(3)}-\tau H_{(3)})\wedge \Omega=
(2\pi)^2\alpha' (f\cdot \Pi-\tau h\cdot \Pi).} The K\"ahler
potential for the dilaton-axion and complex moduli is given by
\eqn\kahler{ K=K_{\tau}+K_{\psi}=-\ln(-i(\tau-\bar{\tau}))-\ln(i\int_{M}\Omega\wedge
\bar{\Omega})= -\ln(-i(\tau-\bar{\tau}))-\ln( - i
\Pi^{\dagger}\cdot\Sigma\cdot \Pi).}

\subsec{Conditions for solutions}

The tadpole condition for the model of interest is given by
\xis.
This provides a restriction on the fluxes we can use, i.e. on our allowed
class of superpotentials -- we must keep $N_{\rm flux} \leq 972$.

We will be searching for nonsupersymmetric flux vacua
which obey the equations
\eqn\condvacua{D_{\tau}W=0\quad{\rm and}\quad  D_{\psi}W=0.}
They are nonsupersymmetric because in the full IIB string theory, they
will generically have $D_{\rho} W \sim W \neq 0$ where $\rho$ is a
K\"ahler modulus.  Note however that in the toy ensemble of
\refs{\AD,\denef}, such vacua are called supersymmetric,
since the K\"ahler moduli are not part of the toy ensemble.

\newsec{Small $\psi$ region}
\subsec{Periods, K\"ahler potential, Metric and Curvature form}
The periods for the small $\psi$, $|\psi|<1$ region are given by
\eqn\persmall{
w_{j}(\psi)=(2\pi i)^3{\pi \over 8}\sum_{n=1}^{\infty}
{1\over \Gamma(n)(\Gamma(1-{n\over 8}))^4\Gamma(1-{n\over 2})}
{\exp\left({7 \pi i\over 8} n\right)\over \sin \left({\pi n\over 8}\right)}
{(4\alpha^{j}\psi)^n\over \psi}
}
where $\alpha=\exp\left({\pi i\over 4}\right)$.
The corresponding period vector is given by
\eqn\perv{
w^{T}=(w_2(\psi),w_1(\psi),w_0(\psi),w_7(\psi))~.
}
We will want the periods in a symplectic basis.  In such
a basis, the period vector
$\Pi^{T}(\psi)=(\CG_1(\psi),\CG_2(\psi),z^1(\psi),z^2(\psi))$
is provided by the linear transformation
\eqn\lintr{\Pi ~=~m \cdot w,} where the matrix $m$ is given by

\eqn\misa{m =\pmatrix{-{1 \over 2} & -{1 \over 2} & {1 \over 2}
& {1 \over 2}\cr 0 & 0 & -1 & 0\cr -1 & 0 & 3 & 2 \cr 0 & 1 & -1 &
0\cr}.}

The period vector in the symplectic basis can be written in the form
\eqn\persympsmall{
\Pi(\psi)=\sum_{n=0}^{\infty}c_{2n}p_{2n}\psi^{2n}.
}
Here
the vectors $p_{2n}$ are given by \eqn\pktilde{p_{2n} = ~m \cdot (\alpha^{2(2n+1)},
\alpha^{(2n+1)}, 1, \alpha^{7(2n+1)})^T}
and the first few constants $c_{2n}$ of interest are given by
\eqn\cis{\eqalign{& c_0 = (2\pi i)^3 {\sqrt{\pi} \over 2 \Gamma^4(7/8)}
{\exp({7 \pi i\over 8
          })\over \sin({\pi \over 8})}, \qquad
 c_2 = -(2\pi i)^3 {2\sqrt{\pi} \over  \Gamma^4(5/8)} {\exp({5 \pi i\over 8
          })\over \sin({3\pi \over 8})}, \cr
& c_4 = (2\pi i)^3 {4\sqrt{\pi} \over  \Gamma^4(3/8)} {\exp({3 \pi i \over 8
          })\over \sin({5\pi \over 8})},\qquad~
c_8=(2\pi i)^3 {\sqrt{\pi} \over  768\Gamma^4(7/8)} {\exp({ 7 \pi i \over 8
          })\over \sin({\pi \over 8})}.}}

This yields the
following K\"ahler potential for the complex structure moduli
space:
\eqn\kahlsmall{
K_{\psi}=-\ln[\lambda_0+\lambda_2|\psi|^4+\lambda_4|\psi|^8
+\lambda_5\psi^8+\bar{\lambda}_5\bar{\psi}^8+O(|\psi|^{12})]}
where the real constants are given by
\eqn\kahlsmalla{\eqalign{
&\lambda_0=-i|c_0|^2p_0^\dagger\cdot\Sigma\cdot p_0,\qquad
\lambda_2=-i|c_2|^2p_2^\dagger\cdot\Sigma\cdot p_2,\cr
&\lambda_4=-i|c_4|^2p_4^\dagger\cdot\Sigma\cdot p_4,\qquad
\lambda_{5}=-i \bar{c}_0 c_8 p_0^\dagger\cdot\Sigma\cdot p_0. }}
Note that  $p_0^\dagger\cdot\Sigma\cdot p_0=2i(\sqrt{2}+2)$
and $p_2^\dagger\cdot\Sigma\cdot p_2=-p_4^\dagger\cdot\Sigma\cdot p_4=2i(\sqrt{2}-2)$.

The resulting K\"ahler metric takes the form
\eqn\metricsmall{
g_{\psi\bar{\psi}}=-4{\lambda_2\over \lambda_0}|\psi|^2\left(
1+\left(4{\lambda_4\over
\lambda_2}-2{\lambda_2\over\lambda_0}\right)|\psi|^4+O(\psi^8)
\right).}
Using the expression for the curvature form
\eqn\curvfrm{
R_{\psi\bar{\psi}}=R^{\psi}_{\psi\bar{\psi}\psi}=-\partial_\psi
\partial_{\bar{\psi}}\ln(g_{\psi\bar{\psi}})}
one finds the following expression of interest
\eqn\curv{
R_{\psi\bar{\psi}}
=-4|\psi|^2\left(4{\lambda_4\over \lambda_2}-2{\lambda_2\over\lambda_0}\right)+O(|\psi|^6)
.}
Note also that one finds
\eqn\Ksmall{
\partial_{\psi}K_{\psi}
=-2{\lambda_2\over\lambda_0}|\psi|^2\bar{\psi}+O(\psi^7),
}
which is of use in evaluating K\"ahler covariantized derivatives
with respect to $\psi$.

\subsec{Distribution of flux vacua}

Given these facts about the Calabi-Yau moduli space geometry, it is now possible to work
out the detailed prediction from \S2.1.
One sees that
\eqn\detsmall{
\int_{\CM}2\pi i\, c_1(\psi)={\rm const}\int_{0}^{r}|\psi|^2 |\psi|d|\psi|
={\rm const}{r^4\over 4}
}
for (an eight-fold cover of) a small piece of the
fundamental region of
complex structure moduli space $\CM$: $|\psi|<r$.

\medskip \ifig\figI{A plot of a numerical evaluation of
$\int_{\CM}2\pi i\, c_1(\psi)$.
The monte carlo simulation data for each point is: number
of random fluxes $N=10^8$;
random flux interval $f,h\in (-40,40)$.
 The data is fit by the curve ${700000\, {r^4\over 4}}$.}
{\epsfxsize=0.8\hsize\epsfbox{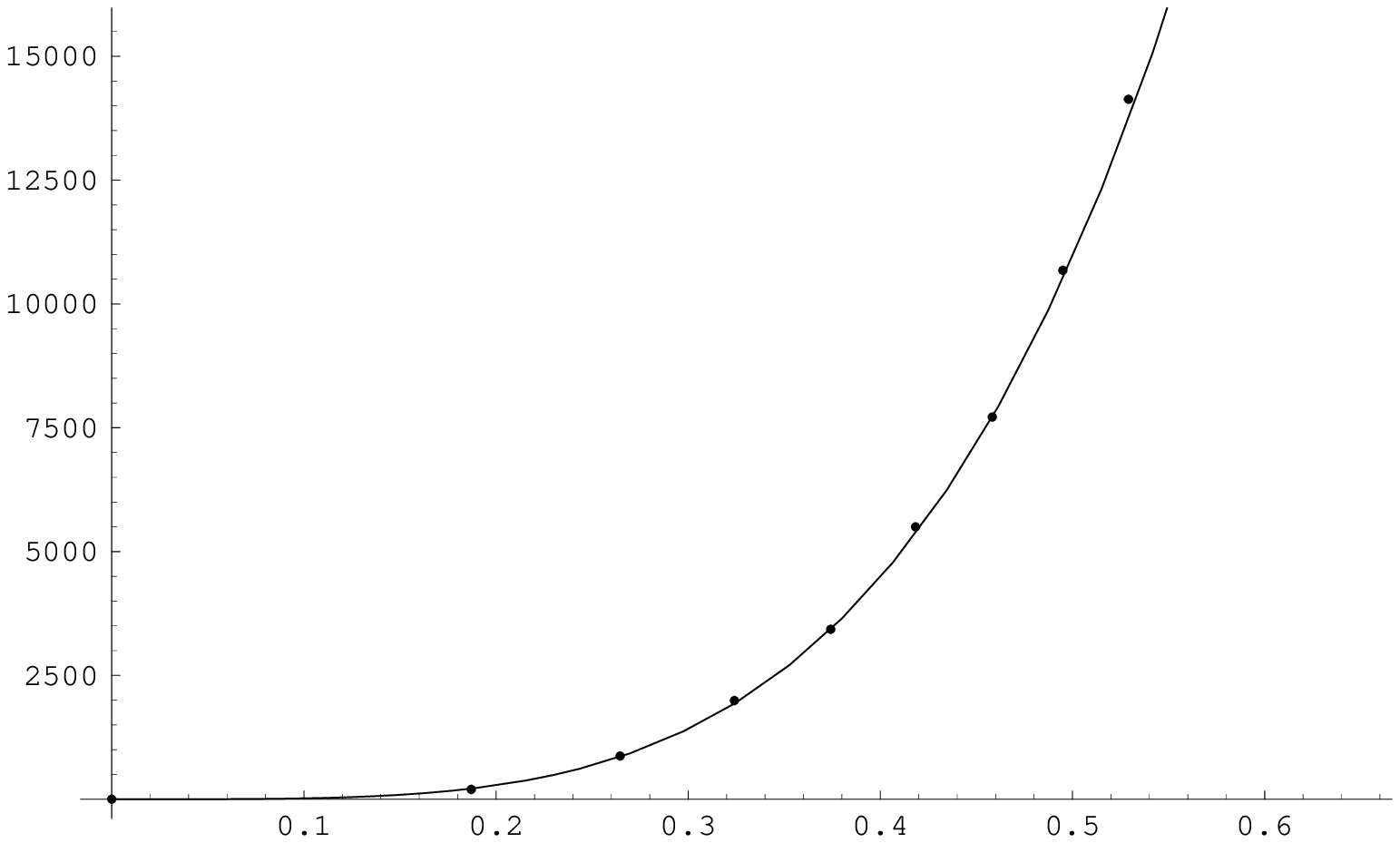}}

The equations for flux vacua, keeping terms up to order $\psi^2$, become
\eqn\condsmalla{\eqalign{
&D_{\tau}W=0\;\;\Rightarrow \;\;(f-\bar{\tau}h)\cdot(c_0
p_0+c_2 p_2 \psi^2)=0;\cr
&D_{\psi}W=0\;\;\Rightarrow \;\;(f-\tau h) \cdot(c_2
p_2+2c_4 p_4 \psi^2-{\lambda_2\over\lambda_0} c_0 p_0
\bar{\psi}^2)=0. }}
This provides two equations
\eqn\condsmallb{
\tau={f\cdot
p_0^{\dagger}+{\bar{c}_2\over \bar{c}_0}f\cdot
p_2^{\dagger}\bar{\psi}^2 \over h\cdot
p_0^{\dagger}+{\bar{c}_2\over \bar{c}_0}h\cdot
p_2^{\dagger}\bar{\psi}^2}
={f\cdot p_2+{2c_4\over
c_2}f\cdot p_4 \psi^2 -{\lambda_2 c_0\over \lambda_0 c_2} f\cdot
p_0 \; \bar{\psi}^2 \over h\cdot p_2+{2 c_4\over c_2}h\cdot p_4
\psi^2 -{\lambda_2 c_0\over \lambda_0 c_2} h\cdot p_0 \;
\bar{\psi}^2}.}

The last one gives a linear equation for $\psi^2$ such that
\eqn\condsmallc{a+b_1\psi^2+b_2\bar{\psi}^2+O(\psi^4)=0,}
where we defined the following quantities depending on fluxes
\eqn\condsmalld{\eqalign{
&a=f\cdot p_0^\dagger\, h\cdot p_2-h\cdot p_0^\dagger\, f\cdot p_2;\cr
&b_1={2c_4\over c_2}(h\cdot p_4\, f\cdot p_0^\dagger-f\cdot p_4\, h\cdot p_0^\dagger );\cr
&b_2={\lambda_2 c_0\over \lambda_0 c_2}
(f\cdot p_0\, h\cdot p_0^\dagger-h\cdot p_0\, f\cdot p_0^\dagger)
+{\bar{c}_2\over \bar{c}_0}
(f\cdot p_2^\dagger\,h\cdot p_2-h\cdot p_2^\dagger\, f\cdot p_2).
}}
Introducing the notation $\psi^2\equiv x+iy$
one gets the following system of linear equations
\eqn\condsmalle{\eqalign{
&{\rm Re}(a)+{\rm Re}(b_1+b_2)x-{\rm Im}(b_1-b_2)y=0;\cr
&{\rm Im}(a)+{\rm Im}(b_1+b_2)x+{\rm Re}(b_1-b_2)y=0. }}

\medskip \ifig\figI{The numerical results for the number of vacua with $N_{\rm flux}<L$
for $L\in(1,972)$.
The monte carlo simulation data is: number of random fluxes $N=10^{9}$;
random flux interval $f,h\in (-100,100)$; complex structure
$\psi$ space region $|\psi|^2<0.5$.
The data is fit by the curve ${L^3\over 183000}$.}
{\epsfxsize=0.8\hsize\epsfbox{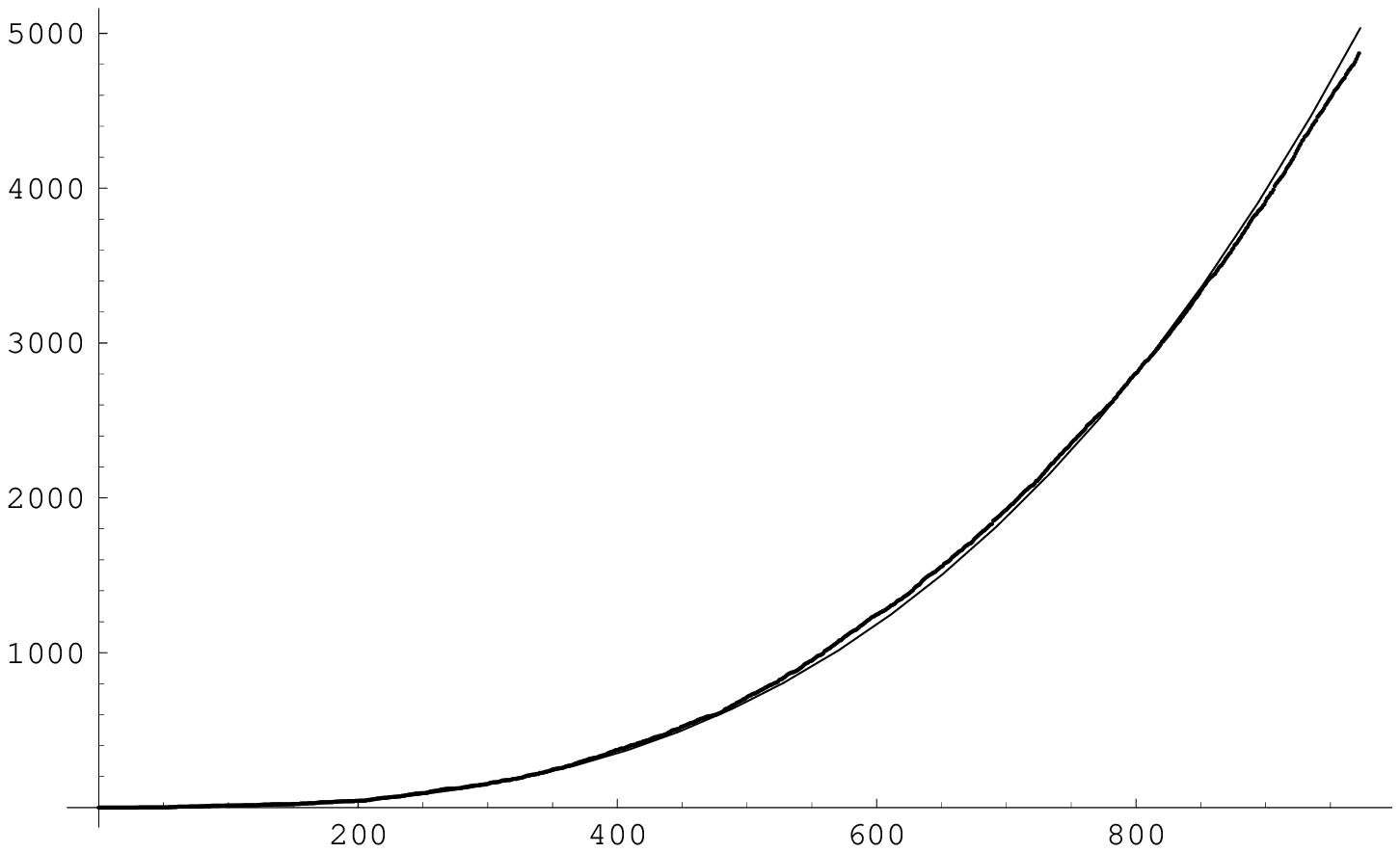}}

We used monte carlo simulations to check the validity of the formula
\detsmall\ and the scaling of the number of vacua with $L$. We chose
correspondingly $10^8$ and $10^{9}$ random fluxes, taking integer values $f,h$
in the ranges indicated in the figure captions.
The number of solutions with $L \le 972$
resulting in small $\psi$ is determined numerically from
eqs. \condsmallb\-\condsmalle\
and is plotted in Fig.1. Fig.2 shows the dependence of the
total number of vacua on $L$.

One sees that the curve in Fig. 1 is in very good agreement with
the prediction \indexd.
It is interesting that in Fig. 2, one sees a rather clean $L^3$ growth of
the number of vacua.  The naive formula \nsol\ would predict
$L^2$, while the more refined formula proved by Ashok and Douglas
under certain assumptions would predict $L^4$.  In fact, as
described in \S2.2\ and the revised version of \denef, the
correct scaling for a small region in $\psi$-space is a
complicated subject.  It may be that exceedingly large $L$
would be required to attain the scaling \ntots, in this
region of moduli space.

\newsec{Conifold region}
\subsec{Periods, K\"ahler potential, Metric and Curvature form}

The periods in a symplectic basis in the vicinity of the conifold point $\psi=1$
can be given to the first order by the following expressions
(here $x\equiv 1-\psi$ and $|x|\ll 1$)
\eqn\perconi{\eqalign{
& \CG_1(x)=(2\pi i)^3[a_0+a_1x+O(x^2)],\cr
& \CG_2(x)={z^2(x)\over 2\pi i}\ln(x)+(2\pi i)^3[b_0+b_1x+O(x^2)],\cr
& z^1(x)=(2\pi i)^3[c_0+c_1 x+O(x^2)],\cr
& z^2(x)=(2\pi i)^3[d_0+d_1 x+O(x^2)].
}}
where the constants can be approximated by the
following numbers
\eqn\perconia{\eqalign{
&a_0=-1.774 i,\qquad\qquad\;\;\, c_0=4.952-5.321i, \cr
&a_1=1.227i, \qquad\qquad\;\;\;\;\;\; c_1=-4.488+3.682i,\cr
&b_0=-1.047, \qquad\qquad\;\;\;\;\; d_0=0,\cr
&b_1=0.451+0.900 i,\quad\quad~ d_1=1.800 i.
}}

The K\"ahler potential for complex structure modulus
is given by \eqn\Kconi{ K_{\psi}=-\ln[
\mu_0+\mu_1 x + \bar{\mu}_1\bar{x} +\mu_2 |x|^2\ln|x|^2
+\mu_3 |x|^2+\mu_4 x^2+\bar{\mu}_4 \bar{x}^2+O(|x|^3\ln|x|)],
} where the relevant constants $\mu_0$, $\mu_1$, $\mu_2$ and $\mu_3$ are
given by
\eqn\Kconia{\eqalign{
& \mu_0=i(2\pi)^6(a_0 \bar{c}_0-c_0 \bar{a}_0),\qquad
 \mu_1=i(2\pi)^6(\bar{c}_0 a_1-c_1 \bar{a}_0-d_1\bar{b}_0),\cr
& \mu_2=(2\pi)^5|d_1|^2,\qquad
  ~~~~~~~~~~~~
  \mu_3=i(2\pi)^6(\bar{c}_1 a_1-\bar{a}_1 c_1+\bar{d}_1 b_1-\bar{b}_1 d_1).
}}
One finds the following expression for the K\"ahler metric
\eqn\metrconi{
g_{x\bar{x}}=-{\mu_2\over\mu_0}\ln|x|^2
+\left({|\mu_1|^2\over \mu_0^2}-{2\mu_2+\mu_3\over \mu_0}\right)+O(|x|\ln|x|).
}
Then the curvature form is
\eqn\Rconi{
R_{x\bar{x}}={1\over 4|x|^2}{1\over (\ln|x|+C)^2},
}
where the constant $C$ is determined to be
\eqn\RicciC{
C=1-{|\mu_1|^2\over 2\mu_0 \mu_2}+{\mu_3\over 2\mu_2}\approx -0.738.}
In computing K\"ahler covariantized derivatives with respect to
$\psi$, it is also useful to note that
\eqn\kahldirconi{
\partial_{x}K_{\psi}=-{\mu_1\over \mu_0}-{\mu_2\over\mu_0}\bar{x}\ln|x|^2+O(x).
}

\subsec{Distribution of flux vacua}

\medskip \ifig\figI{Each point is a vacuum on the $x=1-\psi$ complex plane.
The monte carlo simulation data is: number of random fluxes $N=5 \times 10^7$;
random flux interval $f,h\in (-100,100)$; complex structure
$\psi$ space region $|x|<0.04$. There are 11249 vacua, but
6306 of them arise at $|x|<.00001$ and have been removed from
the plot (they would all cluster at the origin).}
{\epsfxsize=0.8\hsize\epsfbox{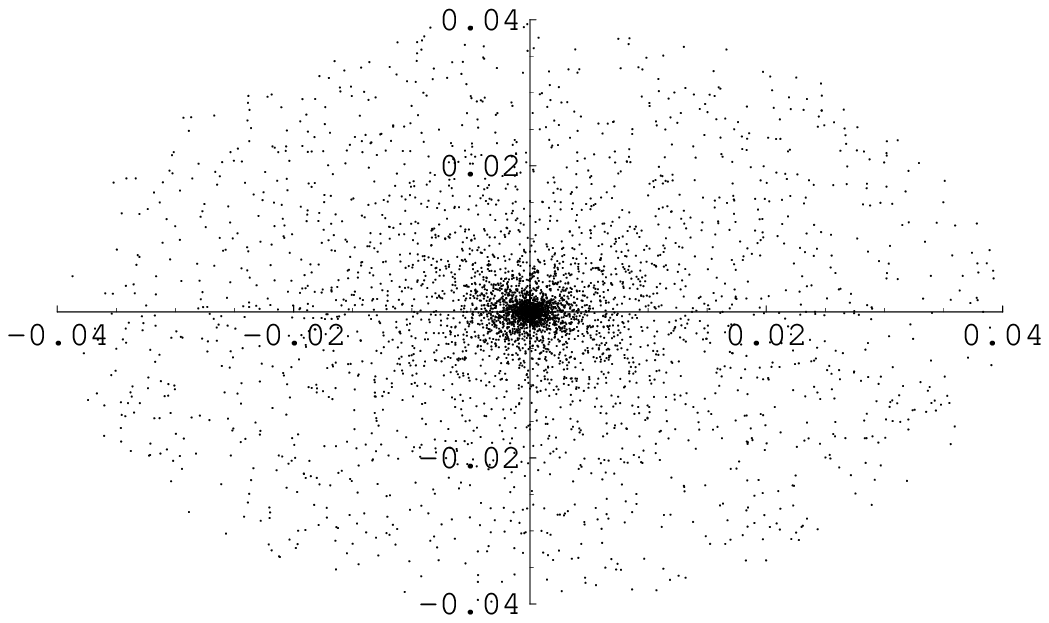}}

\medskip \ifig\figI{The plot of a numerical evaluation of $\int_{\CM}
2\pi i\, c_1(\psi)$.
The monte carlo simulation data for each point is: number
of random fluxes $N=10^7$;
random flux interval $f,h\in (-60,60)$.
The data is fit by the curve $-{200000\over \rho+C}$,
where $\rho=\ln r$.
The conifold point $r=0$ is at $\rho=-\infty$
for this coordinate.} {\epsfxsize=0.8\hsize\epsfbox{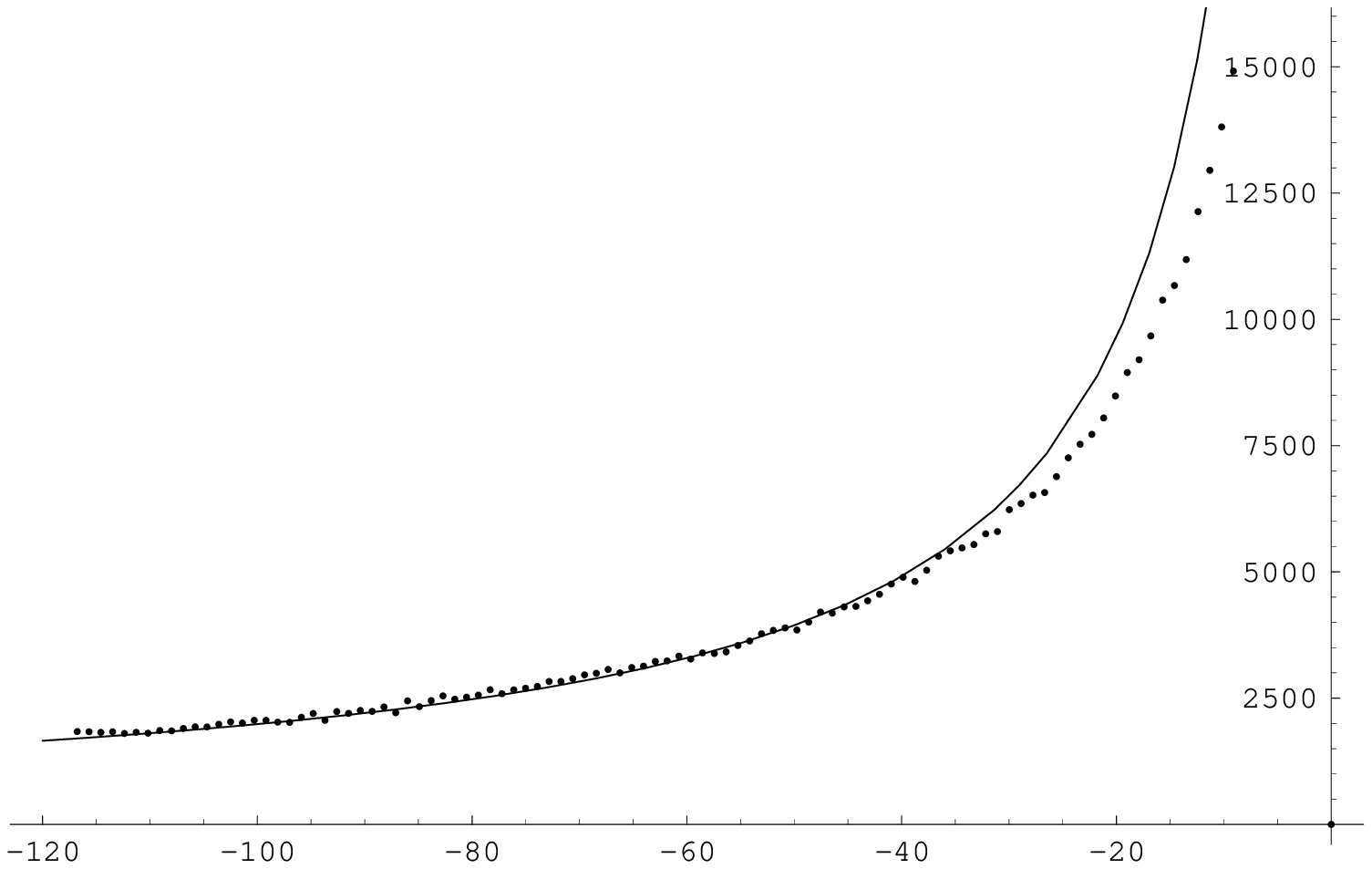}}

\medskip \ifig\figI{The numerical results for the  number of vacua with $N_{\rm flux}<L$
for $L\in(1,972)$.
The monte carlo simulation data is: number of random fluxes $N=10^{10}$;
random flux interval $f,h\in (-200,200)$; complex structure
$\psi$ space region $|x|<0.001$.
The data is fit by the curve ${L^3\over 165000}$.}
{\epsfxsize=0.8\hsize\epsfbox{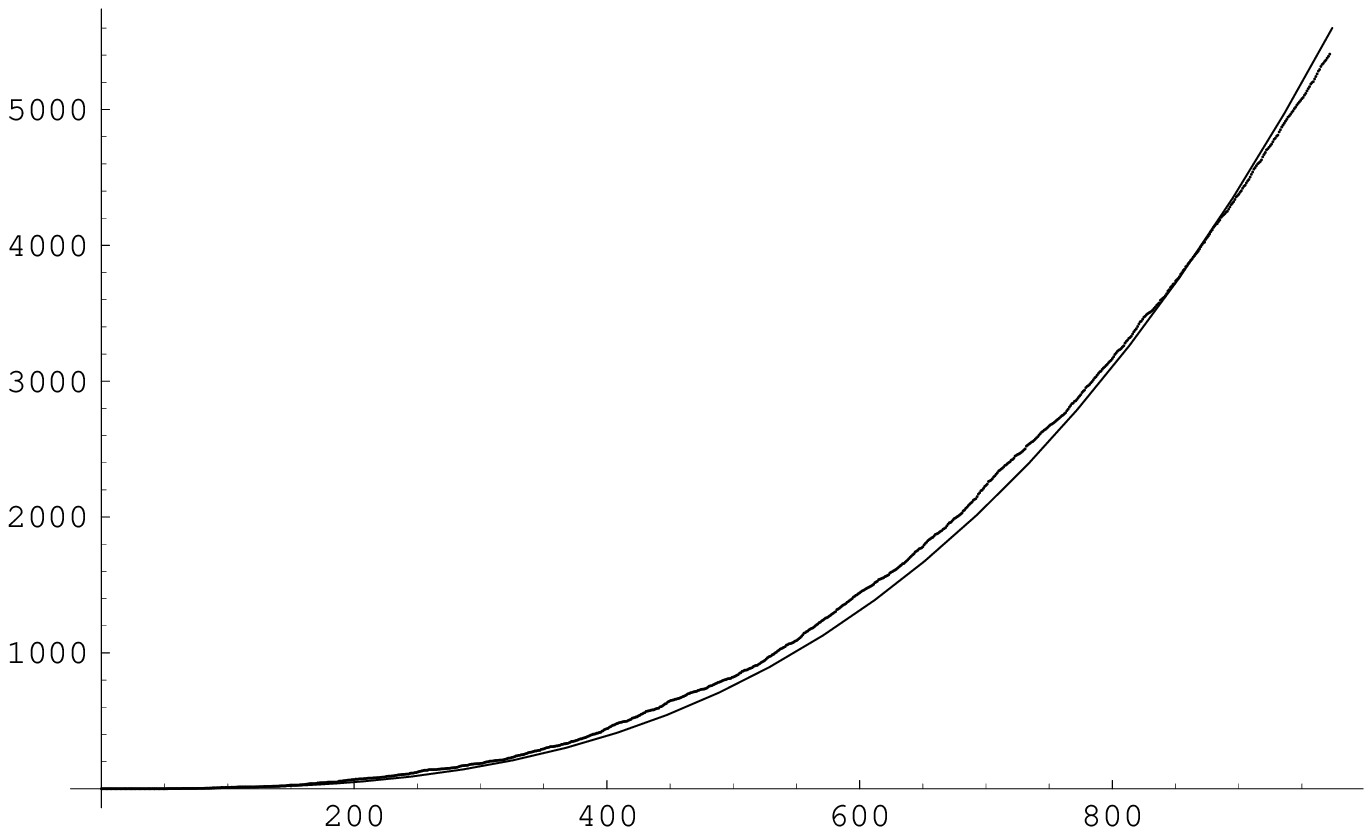}}

We numerically solved for vacua around $\psi=1$.  To compare with the prediction
from \S2.1, we need to determine the curvature and K\"ahler form of the moduli
space.
Using the results of \S5.1, we find
(including only the leading contribution coming from the curvature form)
\eqn\detconi{
\int_{\CM}2\pi i\,c_1(\psi)={\rm const}\int_{0}^{r}{1\over|x|^2}
{1\over(\ln|x|+C)^2} |x|d|x| =-{{\rm const}\over \ln r+C} }
for the
portion of the fundamental region of complex structure moduli space
$\CM$: $|x|<r$.

The equations for flux vacua take the form
\eqn\condconia{
D_{\tau}W=0\quad\Rightarrow\quad
\tau={f\cdot \Pi^{\dagger}\over h\cdot \Pi^{\dagger}}=
{f_1\bar{a}_0+f_2\bar{b}_0+f_3\bar{c}_0\over h_1\bar{a}_0+h_2\bar{b}_0+h_3\bar{c}_0}
+O(|x|\ln|x|);
}
\eqn\condconib{\eqalign{
& D_{\psi}W=0\quad\Rightarrow\quad
 \ln(x)=-{2\pi i\over d_1}\left[
{(f_1-\tau h_1)(a_1-{\mu_1\over \mu_0}a_0)
+(f_2-\tau h_2)(b_1-{\mu_1\over \mu_0}b_0)\over f_2-\tau h_2}+\right.\cr
&\left.\qquad\qquad\qquad\qquad\qquad\qquad
\qquad\qquad +{(f_3-\tau h_3)(c_1-{\mu_1\over \mu_0}c_0)
+(f_4-\tau h_4)d_1
\over f_2-\tau h_2}\right]-1.
}}

We performed a monte carlo simulation, solving the equations of motion for
randomly selected fluxes.
In Figure 3, we plot the distribution of vacua near $\psi=1$ on the $x=1-\psi$
plane.
The most striking thing about Figure 3 is the clustering of vacua around the
conifold point \refs{\denef,\dtalk}.
This would be expected from the curvature dependence of the
result for the index \indexd\ (which, as shown in Figure 4, is in good
agreement with our results).  This is interesting both because it shows that
the conjecture \indexd\ contains much more information than a plausible
boring conjecture like ``vacua are uniformly distributed on the moduli space,''
and that furthermore the detailed distribution we find experimentally
is in fact captured by the
simple formula \indexd!

As described in \refs{\GKP,\KKLT}, the conifold
point is particularly interesting because
it provides a natural mechanism for flux-generated hierarchies
(as in the noncompact Klebanov-Strassler solution \KS), and can provide
small numbers which help with moduli stabilization or attaining small vacuum
energy.  It could therefore also be physically important
that it dominates the distribution
in Figure 3.
Flux vacua near the conifold \GKP\ (see also \Herman)
can explain light scalars by warping (instead of
protecting their masses by supersymmetry).  Therefore,
if this feature persisted in the
full problem of characterizing the string landscape, it could mean that
the correct stringy notion of naturalness allows
warping as a natural explanation
for light scalars (as in Randall-Sundrum scenarios \RS), in
addition to low-energy supersymmetry.
Of course
this is mere speculation at this point, but indicates the kinds of lessons one
could hope to learn.

As shown in Fig. 5, for the range of $L$ most relevant to Model A
of \GKTT, we again seem to obtain close to $L^3$ scaling for the number
of vacua in this region of moduli space.  It is quite possible that
at much larger $L$, this behavior would be modified to $L^4$ in
accord with \ntots, and that a more detailed theory could explain
the $L^3$ scaling visible here.

\newsec{Conclusions}

In this paper, we have seen that very simple conjectures can serve to capture
the detailed structure of the numbers and
distribution of flux vacua on Calabi-Yau moduli spaces.
While the flux potentials are surely not the full story in string theory, they already
encapsulate very rich behavior (including moduli stabilization, warping, supersymmetry
breaking, etc.), and may exhibit much of the complexity
of the full story.  It is therefore heartening that simple conjectures like those
of \S2\ can essentially classify such large ensembles of vacua in simple terms.
This increases our confidence that somewhere down the line, similar
simple conjectures
which capture basic aspects of the structure of the full space of string vacua may become
available.

The most interesting physical feature to emerge is the existence of special
loci in the moduli space which serve as attractors, in the sense that a
large fraction of the flux vacua cluster around them.  In the one-dimensional
moduli space we examined here, the conifold plays this role, in a way
that is immediately apparent from Figure 3.  The existence of such attractors
in a more general setting could be important in helping to define a stringy
notion of naturalness.

A useful generalization of this work would involve the study of
models with many complex structure moduli.
Finding concrete models
which meet the requirements of \KKLT\ or its generalizations
should also become possible,
as we gain the ability to scan over large sets at once.\foot{After
this paper was completed, the interesting work \ddf\ appeared
where explicit examples along the lines of \KKLT\ are presented,
including K\"ahler modulus stabilization.}
The ensemble where one focuses on IIB flux superpotentials could
also be naturally enlarged
to include
generic Non-K\"ahler type II models \IINK, or
classes of vacua where one includes an analysis of nontrivial
D-brane worldvolume superpotentials (see \lerche\ for a recent review).
This could
be important given that D-brane theories in flux backgrounds can give rise
to very large numbers of vacua (see e.g. \refs{\ps,\douglas}).
And finally, it would be valuable to study other well-motivated
toy ensembles which
characterize the sets of supergravities emerging from others limits of
M-theory, like the heterotic
theory.

\medskip
\centerline{\bf{Acknowledgements}}

We are very grateful to S. Trivedi for early collaboration and for
sharing his insights, and to M. Douglas, S. Shenker, E.
Silverstein and L. Susskind for helpful discussions about the
statistics of flux vacua. S.K. would like to thank G. Dvali and
especially S. Dimopoulos for many interesting conversations about
the ``attractive'' nature of the conifold point and possible
physical implications. 
We are also indebted to O. DeWolfe for pointing out a minor error
in the calculations of \S4, which has been remedied in this version
of our paper. 
P.K.T. is very much thankful to the HETG
group at Harvard University, especially to S. Minwalla, for kind
hospitality at Harvard and for providing a lot of encouragement.
He also thanks all the people of India for enthusiastically
supporting research in String Theory. The work of A.G. was
partially supported by {\cyr RFFI} grant 02-01-00695 and INTAS
03-51-6346. The work of S.K. was supported in part by a David and
Lucile Packard Foundation Fellowship for Science and Engineering,
NSF grant PHY-0097915, and the DOE under contract
DE-AC03-76SF00515. The work of P.K.T. was partially supported by
Harvard University and TIFR.

\listrefs
\end